\documentclass[a4paper,11pt]{article}
\pdfoutput=1 

\usepackage{jinstpub} 

\title{\boldmath Multi-Blade detector with VMM3a-ASIC-based readout: installation and commissioning at the reflectometer Amor at PSI}

\author[a,1]{F. Piscitelli,\note{Corresponding author.}}
\author[a]{F. Ghazi Moradi,}
\author[a]{F.S. Alves,}
\author[b]{M.J. Christensen,}
\author[a]{J. Hrivnak,}
\author[a]{A. Johansson,}
\author[a]{K. Fissum,}
\author[a]{C. C. Lai,}
\author[a]{A. Monera Martinez,}
\author[a,c]{D. Pfeiffer,}
\author[a]{E. Shahu,}
\author[d]{J. Stahn}
\author[a]{and P. O. Svensson}

\affiliation[a]{European Spallation Source ERIC (ESS), \\P.O. Box 176, SE-22100 Lund, Sweden}
\affiliation[b]{European Spallation Source ERIC (ESS), \\Ole Maaløes vej 3, 2200 Copenhagen, Denmark}
\affiliation[c]{CERN, CH-1211 Meyrin, Switzerland}
\affiliation[d]{Paul Scherrer Institute, \\Forschungsstrasse 111, 5232 Villigen, Switzerland}

\emailAdd{francesco.piscitelli@ess.eu}

\abstract{The Multi-Blade (MB) Boron-10-based neutron detector is the chosen technology for three instruments at the European Spallation Source (ESS): the two ESS reflectometers, ESTIA and FREIA, and the Test Beam Line. A fourth MB detector has been built, installed and commissioned for the user operation of the reflectometer Amor at PSI (Switzerland). 
Amor can be considered a downscaled version of the ESS reflectometer ESTIA. They are based on the same Selene guide concept, optimized for performing focusing reflectometry on small samples. The experience gained at Amor is invaluable for the future deployment of the MB detector at the ESS. This manuscript describes the MB detector construction and installation at Amor along with the readout electronics chain based on the VMM3a ASIC. The readout chain deployed at Amor is equivalent of that of the ESS, including the readout master module (RMM), event-formation-units (EFUs), Kafka,  FileWriter and live visualisation tools.}


\arxivnumber{2402.08325}

\begin{document}
\maketitle
\flushbottom

\section{Introduction}
The Multi-Blade (MB)~\cite{MIO_HERE,MIO_MB16CRISP_jinst,MIO_MB2014,MIO_MB2017,MIO_MBAMOR,MIO_MBproc,MIO_MBscattsimu,MIO_MyThesis,MIO_ScientificMBcrisp}
 is a Boron-10-based neutron detector conceived to face the challenging requirements, in terms of spatial resolution and counting rate capability, of neutron reflectometry instruments particularly at the European Spallation Source (ESS)~\cite{ESS,ESS2011}. This detector technology has been extensively characterized and tested at neutron reflectometers.  
\\ Three MB detectors will be deployed at the ESS reflectometers (ESTIA~\cite{INSTR_ESTIA,INSTR_ESTIA0,INSTR_ESTIA1,INSTR_ESTIA2} and FREIA~\cite{INSTR_FREIA,INSTR_FREIA2}) and at the ESS Test BeamLine (TBL). 
\\ The MB detector is modular both in terms of detector units (blade assembly) and readout electronics (see figures~\ref{fig1},~\ref{figsect} and~\ref{fig2}); building a larger detector is possible by simply increasing the number of units, the so-called blade assemblies. The ESTIA detector will be 48 units (or cassettes), FREIA 32 and TBL 14. The same modularity applies to the readout electronics. In order to maximize the rate capability of the detector and its readout chain, the electronics channels of each unit are read out individually. Given this high density of channels the readout is realized with the hybrid based on the VMM3a ASIC~\cite{VMM3a_1} which offers 128 readout channels in a compact spatial envelope of about $80 \times 50 \times 20\,\mathrm{mm^3}$. 
\\ The Amor~\cite{INSTR_ESTIA2,AMOR} reflectometer at PSI employs the same concept of the upcoming ESTIA at ESS based on the Selene guides~\cite{INSTR_ESTIA2}. These guides focus the neutron beam at the sample, resulting into a divergent beam at the detector. This feature allows to perform focused reflectometry measurements optimized for small samples~\cite{INSTR_ESTIA2}. By using the full intensity of the beam, measurements can be performed in a shorter time, but, on the other hand, the detector has to cope with the high intensity. The spatial resolution is also a crucial feature of the detector, it allows to calculate the correct angle of the reflected diverging neutron beam at the detector position~\cite{MIO_MBAMOR}. The MB has been specifically designed for this application, particularly its spatial resolution. 
\\ The MB for Amor consists of 14 cassettes, resulting in an active area of $260 \times 140\,\mathrm{mm^2}$ with $3.5 \times 0.5\,\mathrm{mm^2}$ spatial resolution. The MB for Amor has the same number of units of the MB for TBL, only the enclosure (vessel) differs in order to match the instrument requirements in terms of space. 
\\ The readout electronics chain installed at Amor is identical to the one to be deployed at ESS including the front-end assisters (FEA) (for the VMM3a hybrids) and the readout master module (RMM) (see figure~\ref{fig1}). The data is also sent to the event-formation-unit (EFU)~\cite{MB-DMSC} for event processing, streamed to the PSI Kafka~\cite{DMSC_kafka} cluster from where it is converted into Nexus files used for data reduction and analysis~\cite{EFU_Mukai}.
\\ The MB at Amor can be considered a downscaled version of the one for the ESTIA reflectometer at ESS, where the detector technology, the ESS readout electronics and data chains, are also deployed for the user operations.  
\\ This manuscript describes the installation and commission of the MB detector at Amor. The generic details of the detector mechanics and the VMM3a-based readout chain will be also discussed. 

\section{The Multi-Blade detector}
\subsection{Multi-Blade detector technology description}
Many prototypes of the MB have been built and tested over the past 10 years. The proof of concept for this detector technology has been primarily demonstrated with a first test of a prototype at ILL in 2013~\cite{MIO_MB2014,MIO_MyThesis,MIO_MBproc,MIO_HERE}. A detailed characterization of a Multi-Blade detector has been carried out at the Budapest Neutron Centre in 2015~\cite{MIO_MB2017,MIO_MBscattsimu}. In 2017 a reflectometry demonstrator has been installed and tested at the neutron reflectometer CRISP at the ISIS neutron and muon source in the UK~\cite{MIO_ScientificMBcrisp,MIO_MB16CRISP_jinst}. A full detector characterization was performed along with specular and off-specular reflectivity measurements on several samples. Moreover, background studies, in particular the response of this detector to fast neutrons~\cite{MIO_fastn,MIO_fastnhe3giac,MIO_He3He4fastN} and gamma-rays~\cite{MIO_MB2017}, has been carried out at the Source Testing Facility (STF) at the Lund University in Sweden. During 2019-2020, a Multi-Blade demonstrator has been tested for an entire year at the Amor reflectometer at PSI (CH), serving the instrument as the main detector~\cite{MIO_MBAMOR}.
Table~\ref{tab1} summarizes the characteristics of the MB detector. The intrinsic spurious scattering is defined as the fraction of neutrons that scatters in the detector materials (e.g. detector entrance window, blade substrate, etc.) and get detected as spurious events in other area of the detector where no neutron beam is directed. 

\begin{table}[htbp]
\centering
\caption{\label{tab1} \footnotesize Detector performance~\cite{MIO_HERE,MIO_MB16CRISP_jinst,MIO_MB2014,MIO_MB2017,MIO_MBAMOR,MIO_MBproc,MIO_MBscattsimu,MIO_MyThesis,MIO_ScientificMBcrisp}.}
\smallskip
\begin{tabular}{|l|l|}
\hline
\hline
parameter  &  \\        
\hline
\hline
detection efficiency  &  $45\,\%$ (at 2.5\,\AA) \\
	      &  $57\,\%$ (at 4\AA) \\
	  	  &  $82\,\%$ (at 12\AA) \\
\hline
spatial resolution  &  $0.5 \times 3 \,\mathrm{mm^2}$ \\
\hline
counting rate capability & $>3.5\,\mathrm{kHz/mm^2}$ \\
 & $>60\,\mathrm{kHz} / 30\,\mathrm{mm^2}$ \\
\hline
gamma-ray sensitivity & $<10^{-7}$ \\
\hline
fast neutron sensitivity ($1\,\mathrm{MeV} - 10\,\mathrm{MeV}$)  & $<10^{-5}$ \\
\hline
intrinsic spurious scattering  & $10^{-4}$ \\
\hline
uniformity  & $1.5\,\%$ \\
 \hline
 stability  & $\pm\,2\,\%$ (over days) \\
 \hline
 \hline
\end{tabular}
\end{table}

\begin{figure}[!ht]
\centering
\includegraphics[width=14cm,keepaspectratio]{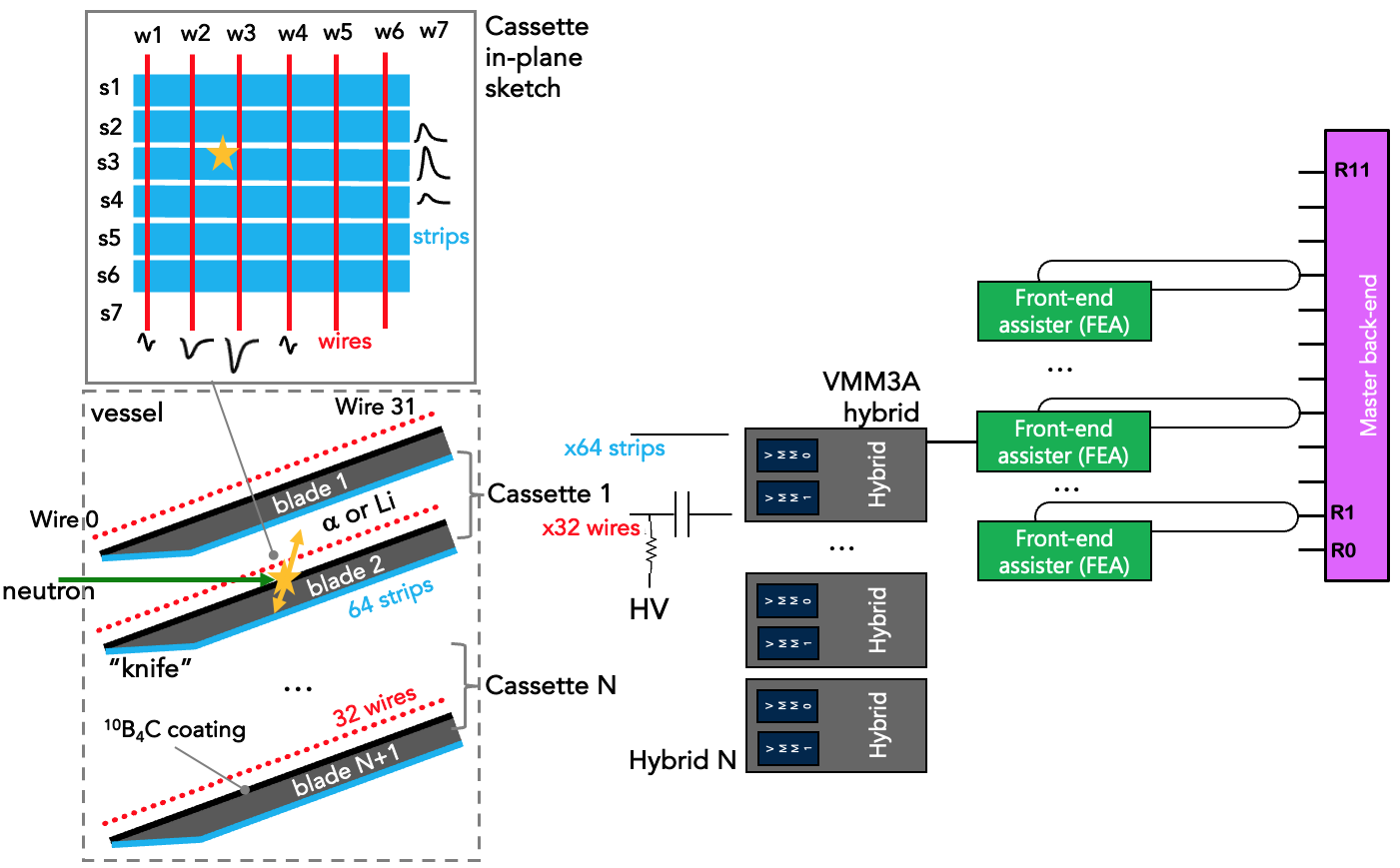}
\caption{\footnotesize Schematic drawing of detector stack and the readout chain.}
\label{fig1}
\end{figure}

The fundamental unit of the MB detector, the blade assembly, consists of a stainless-steel blade coated with enriched Boron carbide. The detector is operated with Ar/CO2 gas mixture (80/20) in a continuous flow. Each blade is inclined at 5 degrees with respect to the incoming neutron beam, this results into an enhanced spatial resolution and rate capability. Boron-10 captures neutrons emitting  a pair of charged Lithium and alpha particles back to back. One side of the blade assembly is covered with 64 strips, it acts as a segmented cathode. On the opposite side, 32 gold-coated tungsten wires (and an extra guard wire) are stretched over the blade by a holding frame (U-frame), perpendicularly to the strips. A “cassette” consisting of 96 channels is defined by two adjacent blade assemblies that delimit a gas volume where the neutron capture fragments are detected. Each cassette is an independent Multi Wire Proportional Chamber (MWPC). 
\\ The blades are sputter-coated with about $8\,\mathrm{\mu m}$ of 95\% $\mathrm{^{10}B}$-enriched boron carbide at ESS Detector Coatings Workshop. It has been shown in~\cite{MIO_MBAMOR,MIO_MB16CRISP_jinst} that this coating thickness is needed to attenuate the neutrons, that can cause scattering within the detector, below $\approx 10^{-4}$-$10^{-5}$~\cite{MIO_MBAMOR}. A thinner coating would allow scattered neutrons to reach the blade, which causes an increased background in the detector. Hence, the coating on the blades acts a as shielding: anything behind the Boron-layer is shielded from the neutron path. For the same reason, the blades are machined with an edge at the front of 2.5 degrees for a length of $45.8\,\mathrm{mm}$ (the “knife”, see figure~\ref{fig1}). The overall blade length is about $130\,\mathrm{mm}$ and covered with wires at $4\,\mathrm{mm}$ pitch. The strip PCB, at the back of the blade, ends $1\,\mathrm{mm}$ away from the edge, in order to avoid spurious scattering caused by neutron beam. Due to the knife shape of a blade, the physical gap between two adjacent blade assemblies (the detecting volume, i.e.~a cassette) is changing across the wire chamber. Wires 0 to 10 are affected by the gap increase assuming wire 0 being the wire at the front of the blade, i.e.~at the edge of the knife. In order to compensate for this physical gap and to keep the gas gain identical across the wires, the voltage applied to these wires is adjusted in steps of about 15~V as the gap increases. Each blade has two high voltage (HV) channels which applies different voltages to the wires 0 to 10 covering the knife side (via a voltage divider) and to the wires 11 to 32 covering flat side of the blade (constant voltage). 
\\ It has been shown in~\cite{MIO_MBAMOR}, that a thin Aluminium foil as detector entrance window decreases the detector background. The Amor detector has a interchangeable entrance window of $100\,\mu\mathrm{m}$. It is interfaced to a flight tube filled with argon to minimize scattering in the $4\,\mathrm{m}$ flight path from the sample to the detector. The latter is also closed with two thin Aluminium foil windows.

The cassette's channels are physically connected to one VMM3a hybrid which has 128 channels. To maintain the modularity of the system 96 out of 128 available channels are utilised in this configuration. Thus one, and only one, VMM hybrid is assigned to each cassette. By keeping channels local to a single VMM hybrid would in principle allow to implement data clustering and reduction per cassette in order to scale the performance. At the moment this processing is done in CPU software, but could be implemented in firmware on the FPGA of each hybrid for even better rate performance. Internally to the VMM3a ASIC, the pre-amplified signal is fed into an ADC (analogue-to-digital converter). Once a signal crosses a set threshold the time-stamp and the ADC peak value is calculated and sent downstream along with the front-end and the channel IDs. The time-stamp is used to calculate Time-of-Flight (ToF) of a neutron event. The triggered channels are then used, along with the ADC values, to calculate the weighted position of the neutron detection in the Event Formation Unit (EFU)~\cite{MB-DMSC}. Moreover, the ADC values are used to perform a further filtering (software threshold) at the software stage to discriminate against background events (i.e. gamma-rays).

\subsection{The Multi-Blade detector for instruments}
\begin{figure}[!ht]
\centering
\includegraphics[width=14cm,keepaspectratio]{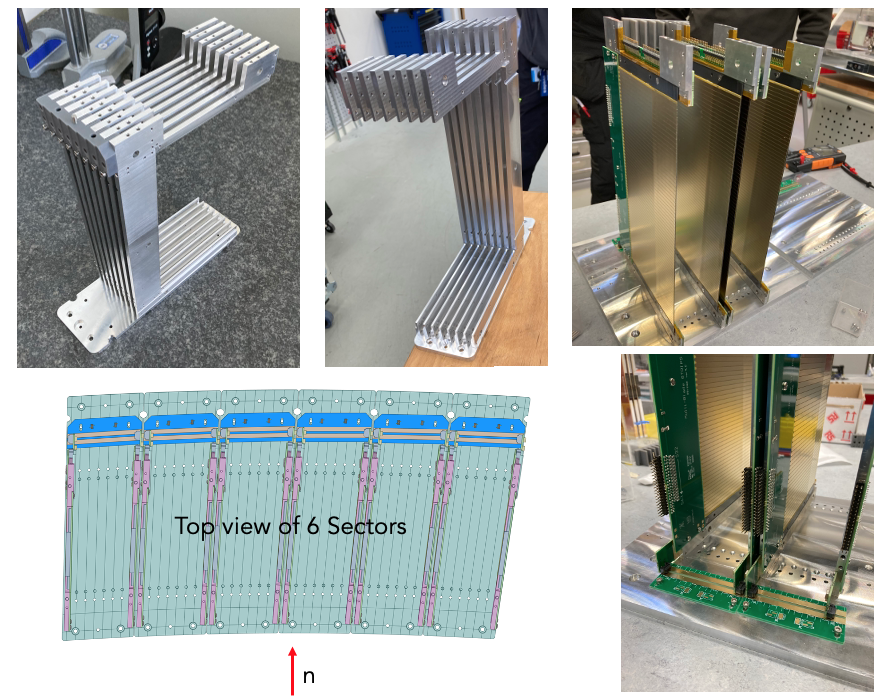}
\caption{\footnotesize One sector of eight cassettes (top left). Top view of six sectors with direction of incoming neutrons (bottom left). A few blade assemblies mounted on the sector or array plate to verify the HV connectivity (right).}
\label{figsect}
\end{figure}

\begin{figure}[!ht]
\centering
\includegraphics[width=14cm,keepaspectratio]{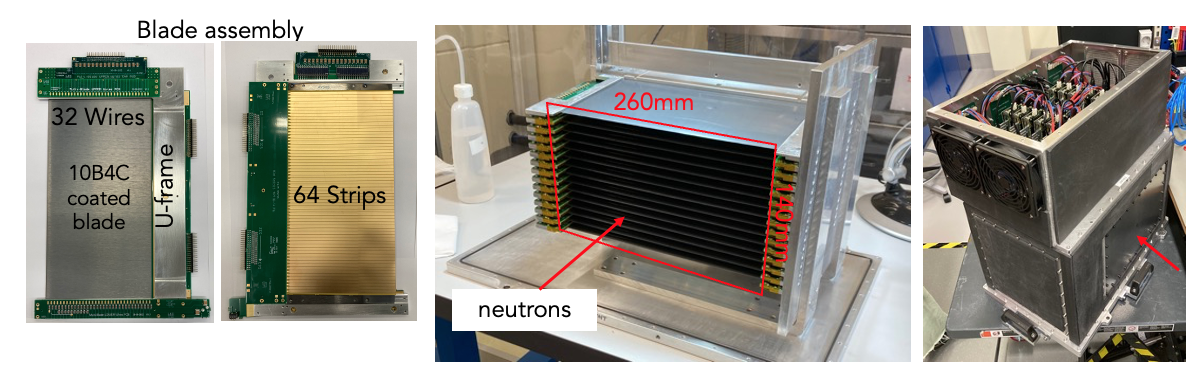}
\caption{\footnotesize The front and the back sides of a blade (left). A stack of 15 blades (14 cassettes) (centre). The MB Amor detector assembled including the electronics box containing the VMM3a hybrids at the top (right).}
\label{fig2}
\end{figure}

\begin{figure}[!ht]
\centering
\includegraphics[width=14cm,keepaspectratio]{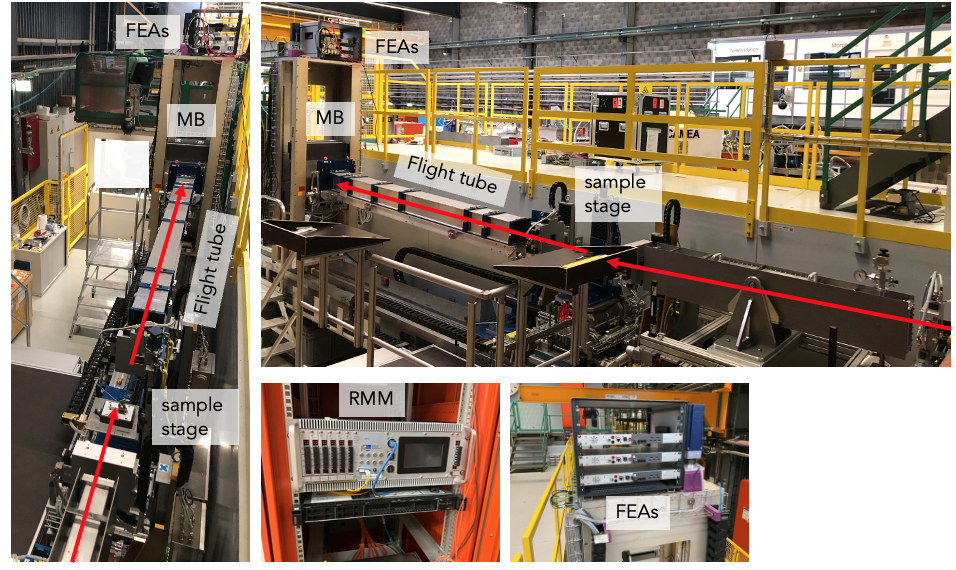}
\caption{\footnotesize Pictures of the installation at Amor showing the neutron path at the reflectometer in red.}
\label{fig3}
\end{figure}

The MB deployed at Amor is the first of four detectors that will be installed permanently on a neutron instrument. A total of 140 blade assemblies have been produced in series and stored. This quantity is enough for the four detectors including spares, given that 49 blades are needed for ESTIA, 33 for FREIA and 15 for TBL and Amor. 
In order to simplify assembly and maintenance, the blades are grouped in sectors. A sector is electrically and mechanically an independent detector (see figure~\ref{figsect}). 
\\ Each sector of 8 blades (7 blades for Amor or TBL) shares the same HV channels and can be powered independently from other sectors. Each HV channel is filtered with a low pass filter at the SHV connectors and an additional filter is present on each blade for both channels. The HV is supplied (see figure~\ref{figsect}) via a spring-loaded connector pressed against an HV strip PCB to simplify installation and maintenance. 
\\ Figure~\ref{fig2} shows the Amor detector comprised of 15 blades (14 cassettes) with the VMM3a electronics box at the top of the gas vessel. 
\\ Figure~\ref{fig3} shows the MB Amor detector installed at Amor. The sample stage and the flight tube are also shown, along with the readout system components. 

\section{Readout system and data chain}
\subsection{Readout topology}
\begin{figure}[!ht]
\centering
\includegraphics[width=10cm,keepaspectratio]{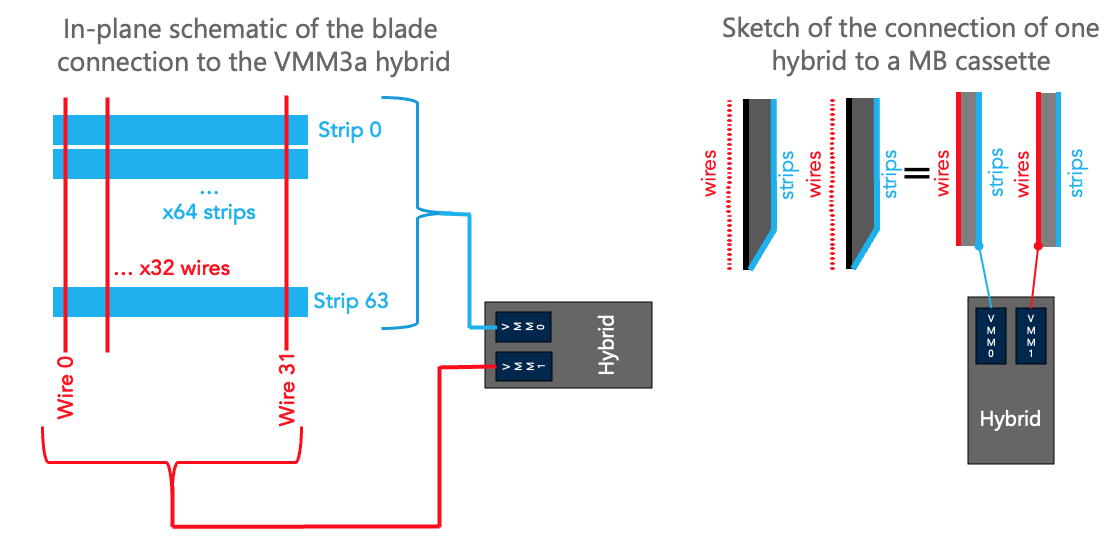}
\caption{\footnotesize Schematic drawing of the connection of one VMM3a hybrid to the wires and strips of a blade.}
\label{figblhyb}
\end{figure}

\begin{figure}[!ht]
\centering
\includegraphics[width=14cm,keepaspectratio]{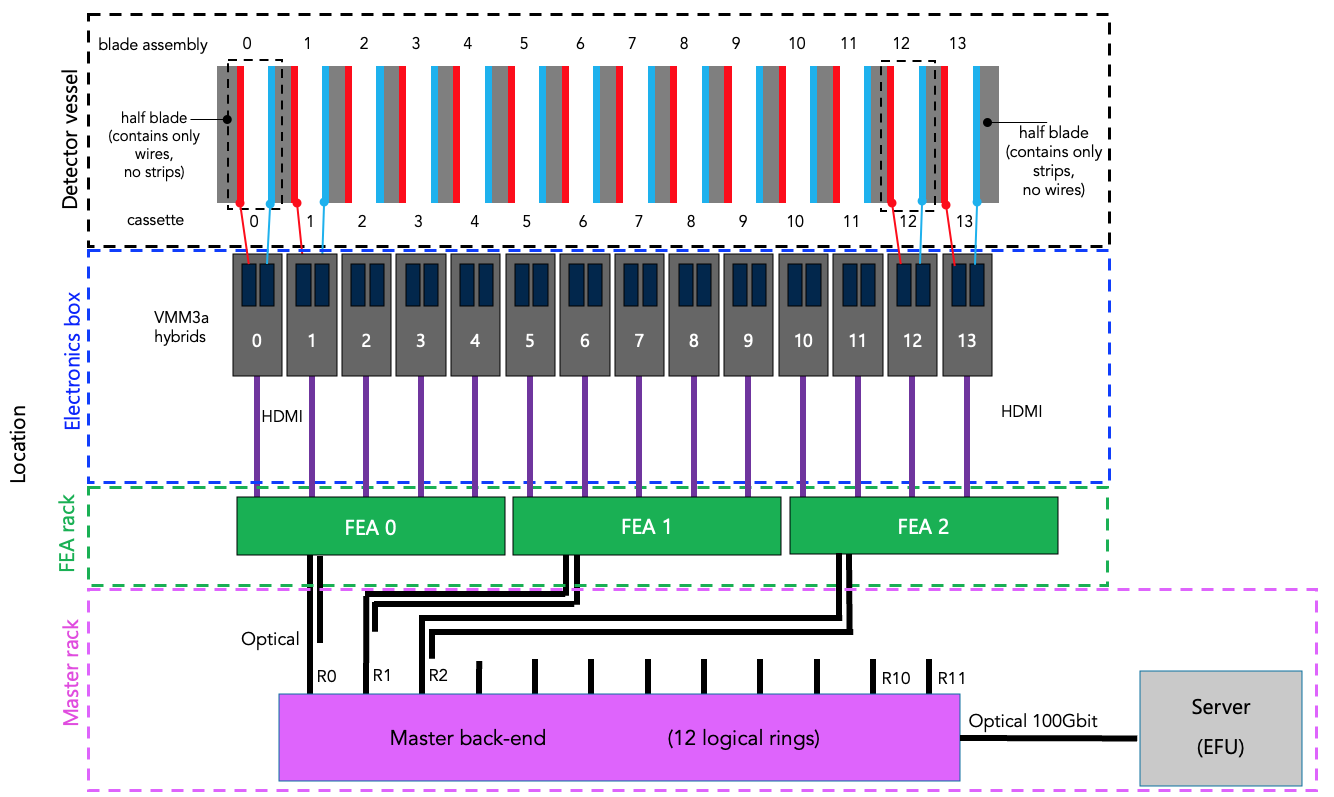}
\caption{\footnotesize Schematic drawing of detector readout system in the case of a 14-cassette detector. }
\label{figarchi}
\end{figure}

As shown in figure~\ref{fig1}, a readout master back-end module (RMM) is the concentrator of the data coming from all the front-end electronics. Each of the ESS instruments will have a RMM which has 12 optical fibre ring interfaces and is compatible with the particular front-end used for a specific detector technology at ESS~\cite{ESS-readout}. The RMM sends the data downstream to the event-formation-unit (EFU)~\cite{MB-DMSC}. The EFU is a software-based data acquisition and processing system that clusters the data and generates a pixel-ID per neutron consisting the strip/wire position for each cassette and the time information. For the MB detector the front-end is a VMM3a hybrid with an Application-Specific Integrated Circuit (ASIC). The hybrid comes with a micro-HDMI LVDS interface for the data transfer. In the present VMM3a firmware version, the hybrid reads out the triggered channel, the time-stamp and the charge information from the VMM3a. This information is transferred from the hybrid to a front-end assister (FEA) consisting of a KC705 FPGA board with SFP+ and HDMI mezzanine interfaces. The FEA interfaces the hybrid HDMI and the optical fibre connection to the RMM. 
\\ The RMM propagates the global timing to all front-end rings and receives the data coming from each node in the ring. Each ring can have multiple nodes, i.e. multiple FEAs. 
\\ Figure~\ref{figblhyb} shows the schematic connection of one hybrid to the electrodes (wires and strips) of a single blade. Figure~\ref{figarchi} shows the architecture of the readout system specifically for a 14-cassette MB detector. Each cassette is readout by one VMM3a hybrid. Up to 5 hybrids are connected to a single FEA which is then synchronized with the RMM via an optical link. The data is concentrated at the RMM and sent to a local server for clustering the readouts into neutron events by the EFU. 
\\ The FEA deployed at Amor can read out up to five hybrids via a HDMI cable and send data to the master module via optical fibre in the readout rings. Each FEA is equipped as well with a $24\,V$ power supply to provide the voltage to the LV distribution boards on the detector electronics box to power the hybrids. Since the MB Amor detector is made up of 14 cassettes, 3 FEAs are necessary to read out the corresponding 14 VMM3a hybrids. Each FEA is a single node in a single ring. Figure~\ref{fig3} shows the installed FEAs in a rack at the top of the detector tower at Amor. 
\bigskip
\subsection{VMM3a ASIC}
The VMM3a~\cite{VMM3a_1} is a custom ASIC developed at Brookhaven National Laboratory. It was designed as the front-end ASIC for Micro-Pattern Gaseous Detectors (MPGD), and is fabricated in the $130\,\mathrm{nm}$ Global Foundries 8RF-DM process (former IBM 8RF-DM). The 64 channels with highly configurable parameters can digitize signals with positive and negative polarity, and meet the processing needs for MB detector signals.
\\ Following is the list of characteristics of the VMM3a ASIC:
\begin{itemize}
  \item $130\,\mathrm{nm}$ CMOS technology
  \item 64 self-triggered input channels
  \item Positive and negative polarity sensitive
  \item Digital block with neighbouring logic, FIFO, multiplexer
  \item Gain $0.5$ to $16\,\mathrm{V/pC}$
  \item Shaping time from $25$ to $200\,\mathrm{ns}$
  \item Input capacitance from few $\mathrm{pF}$ to $1\,\mathrm{nF}$
  \item Time resolution $0.2\,\mathrm{ns}$
  \item Max hit rate per channel $4\,\mathrm{MHz}$
  \item 38 bit per hit (ch, time, charge)
  \item Precision (10-bit) amplitude 
  \item Multiplexed analogue amplitude and timing outputs can be valuable in development and debugging. 
\end{itemize}
Thus the VMM3a represents a valid readout front-end option for the MB. The positive and negative polarities are needed to do the readout of the positive strip signals and the negative wire signals. The signal formation time of the MB detector is at most $300\,\mathrm{ns}$ (on average $150\,\mathrm{ns}$) ~\cite{MIO_MB2017}. This implies that the largest shaping time from the VMM3a must be used ($200\,\mathrm{ns}$) which is sufficient to integrate the signal. As shown in figure~\ref{phsfig}, the Pulse Height Spectrum (PHS) from the Boron-10 neutron capture is clearly showing the energies of the two captured fragments. Strip capacitance in the MB detector is below $50\,\mathrm{pF}$, well below the maximum accepted capacitance by the VMM3a ASIC. For the quality check and debug purposes each channel of the VMM3a can be pulsed internally  with a selectable charge injected through an internal capacitor of $300\,\mathrm{fF}$.  

\subsection{Electronics enclosure, hybrid adapters, low voltage distribution and grounding}
\begin{figure}[!ht]
\centering
\includegraphics[width=14cm,keepaspectratio]{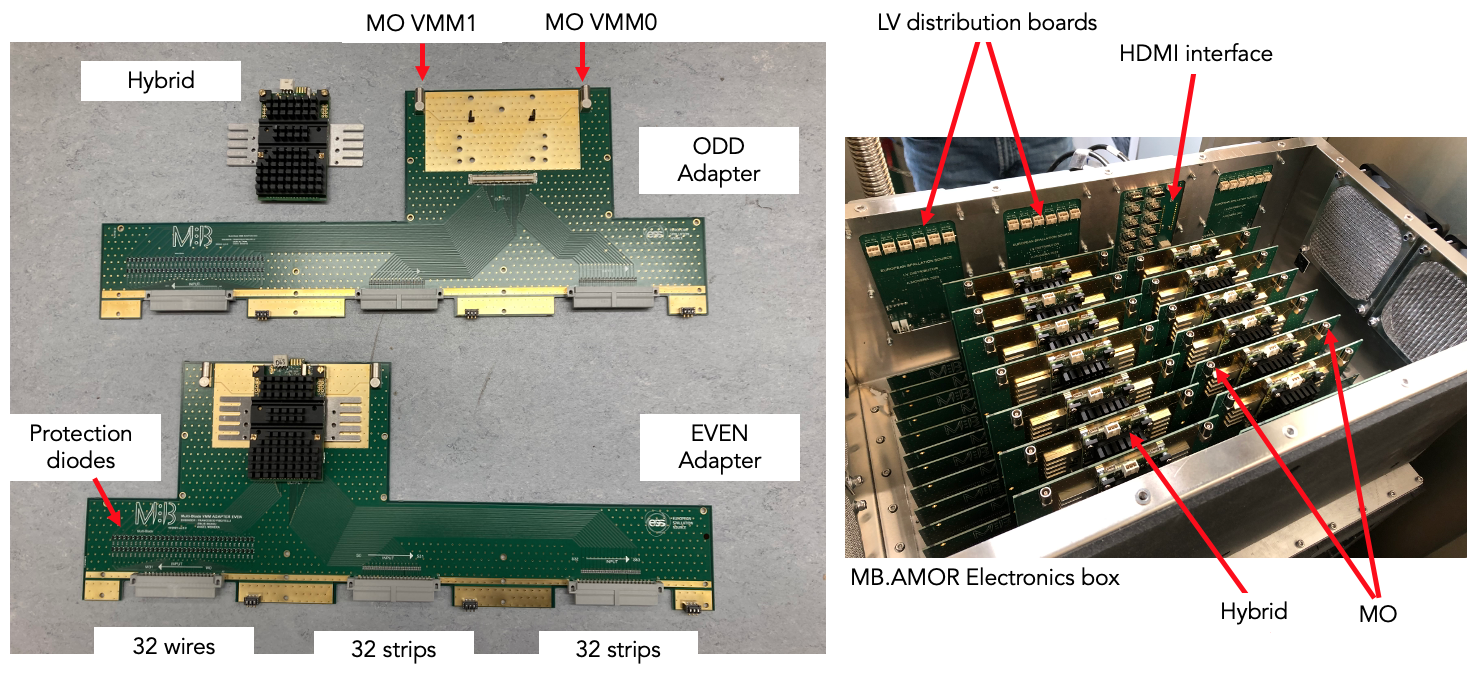}
\caption{\footnotesize The VMM3a hybrid adapters to connect one cassette to one hybrid which are made in two types ODD and EVEN to minimize the space required to stack the hybrids on the detector (left). The 14 hybrids with their respective adapters in the MB Amor electronics box (right). Each adapter has two LEMO connectors to allow for analogue monitoring of the signals.}
\label{figadapt}
\end{figure}

\begin{figure}[!ht]
\centering
\includegraphics[width=10cm,keepaspectratio]{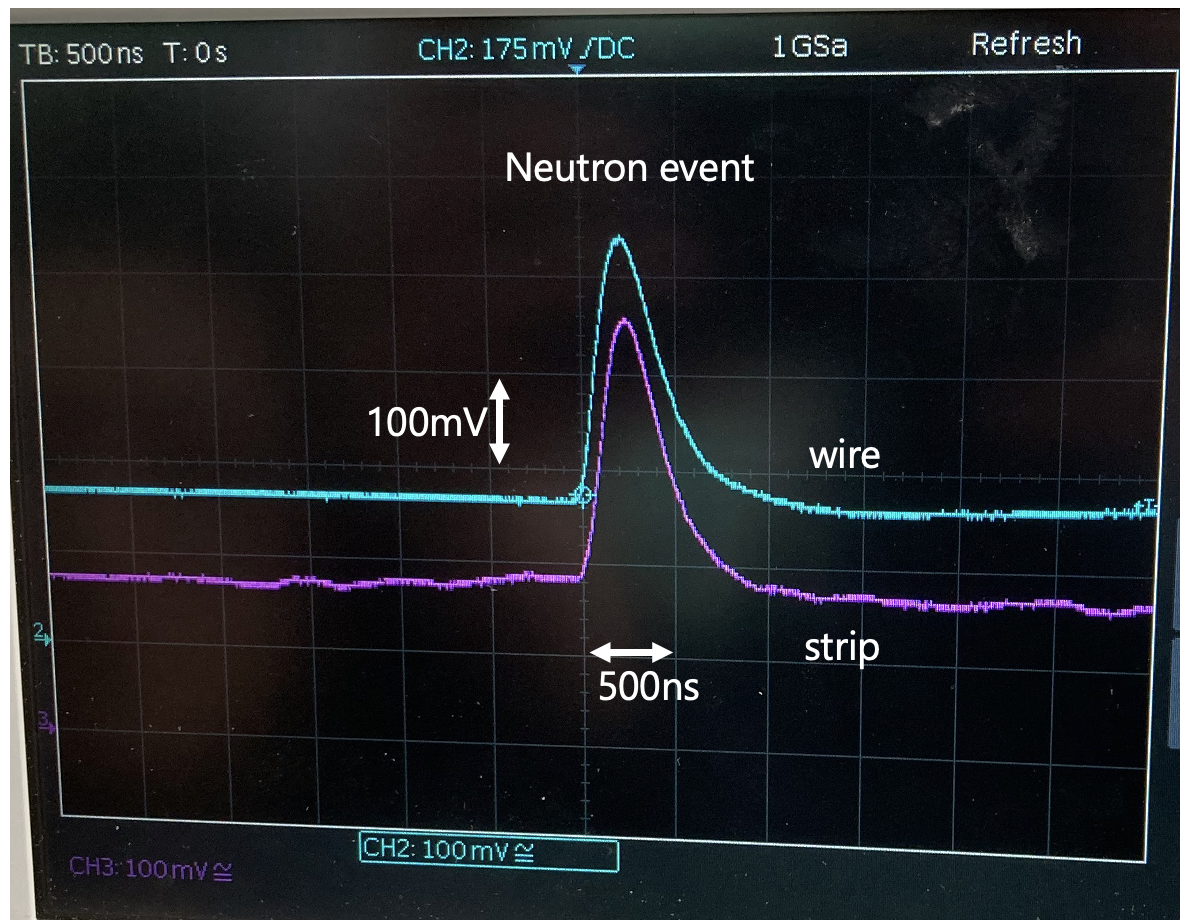}
\caption{\footnotesize A screenshot of the oscilloscope connected to the monitor output (MO) of VMM3a for one cassette, reading the selected strips and wires. The two signals in coincidence are generated by a neutron event.}
\label{figsig}
\end{figure}

A RD51 hybrid~\cite{Lupberger2018,Pfeiffer2022} hosts one Spartan 7 FPGA and two VMM3a ASICs for a total of 128 channels. One hybrid reads one cassette of the MB, hence only 96 channels are used to maintain the modularity of the system. Referring to figure~\ref{figadapt}, the 96 channels of a cassette are connected to a single hybrid through an adapter. The connectors on the MB feed-through have a pitch of $16\,\mathrm{mm}$ each. Since the hybrids are $20\,\mathrm{mm}$ thick each, the adapter was designed in an even-odd configuration and by alternating these two types, inside the electronics box of the detector, the hybrids fit in the space available (see figure ~\ref{figadapt}). The adapter also provides extra circuitry between the detector and the hybrids. In addition to the hybrid HIROSE connector with 128 channels, the adapter also connects the Monitor Output (MO) pins of the hybrids with 2 LEMO connectors. The MO can be programmed via I2C to output several VMM3a internal voltage levels such as the threshold, the pedestal and the voltage level for a selected channel or the values of the internal temperature sensor. This feature is crucial for debugging purposes. Figure~\ref{figsig} shows the analogue wire and strip MO signals in presence of neutron irradiation for one cassette. The strip signal has a higher noise with respect to the wire due to the higher intrinsic strip capacitance connected to the input ($35$ to $50\,\mathrm{pF}$).
\\ The Wire channels are protected via a double layer of protection diodes (BAV99) whereas the strip channels go to the hybrid directly via a $10\,\mathrm{\Omega}$ resistor. Without the additional protection circuit on the adapter, damaged wire channels have been observed. The ESD protection on the hybrid itself (NUP4114) is sufficient for MPGDs and MB strips, but not for MB wires.
\\ One hybrid requires two LV power supplies at $2\,\mathrm{V}$ ($2\,\mathrm{A}$) and $3.3\,\mathrm{V}$ ($<400\,\mathrm{mA}$) for a total of about $6\,\mathrm{W}$. Given the large number of hybrids, delivering these voltages and currents at the detector is not practical. Therefore a LV distribution board, powered at $24\,\mathrm{V}$ ($72\,\mathrm{W}$) from the FEA has been devised to provide the correct power for up to 6 hybrids. It is placed at the detector electronics box close to the hybrids (see figure~\ref{figadapt}). In order to keep the modularity only the 5 hybrids that are read out by the same FEA are powered with the same LV distribution board.
\\ The hybrids are connected to the FEAs via HDMI cables (one cable per hybrid) carrying the LVDS signals and clocks. The HDMI cables are protected with a grounded shield so that the common mode noise interfering with the ground is blocked from entering into the electronics box. This is realized with the HDMI interface PCB (see figure~\ref{figadapt}) that carries the LVDS signals inside the box while connecting the grounds of the cables (inside and outside) with a $1\,\mathrm{m\Omega}$ resistor. The HDMI interface PCB is also connected to the electronics box (which is grounded along with the detector vessel) via the supporting screws. 

\subsection{Rate calculations}
The hardware limitations in terms of rate are the following: 
\begin{itemize}
  \item Per VMM channel $4\,\mathrm{Mhit/s}$ (one hit is 38 bits)
  \item Per hybrid ~\cite{Pfeiffer2022}: $17.6\,\mathrm{Mhit/s}$, i.e. if all channels triggered at the same time, $140\,\mathrm{kHz}$ per channel (one hit is 38 bits)
  \item Per assister (FEA): $31.25\,\mathrm{Mhit/s}$ (32 bits are transferred with $125\,\mathrm{MHz}$ clock or $4\,\mathrm{Gbit/s}$, one hit is 128 bits)
  \item Per ring: $82.5\,\mathrm{Mhit/s}$ (rings are $2 \times 6.6\,\mathrm{Gbit/s}$, one hit is 160 bits)
\end{itemize}
When passing from VMM3a to the Readout Master Module (RMM), the size of the data packet corresponding to a single hit increases. This is due to the addition of VMM, assister and ring identifiers, higher order timestamps and a change of data format for the different stages. The maximum bandwidth per hybrid with two VMM3a is $17.6\,\mathrm{Mhit/s}$ if all channels of both ASICs are triggered simultaneously. One FEA can transfer $31.25\,\mathrm{Mhit/s}$ over the ring to the RMM, which is a little less than the data of two hybrids. One ring on the other hand can transmit $82.5\,\mathrm{Mhit/s}$, or the equivalent of almost 5 hybrids. In the present configuration the hypothetical hardware bottleneck is thus first the FEA and only secondly the ring.
\\ Referring to~\cite{MIO_MB16CRISP_jinst} (page~7, figure~5), the most likely thermal neutron event multiplicity is an event firing one wire and two strips; i.e. 3 hits per neutron event are expected on average. Thus the $82.5\,\mathrm{Mhit/s}$ that a FEA can handle corresponds to about $27\,\mathrm{M events/s}$ in terms of neutron events. If the 5 hybrids, readout by a single FEA, are triggered at the same rate, the maximum rate per hybrid is about $5.5\,\mathrm{M events/s}$ at most. On the other hand, if only one hybrid in one FEA is triggered, the rate is limited to about $17.6\,\mathrm{Mhit/s}$ which corresponds to $5.8\,\mathrm{M events/s}$ on average. In either case the rate limitation of single hybrid is about $5\,\mathrm{M events/s}$, this corresponds to the theoretical limit of a cassette in the MB detector. 
One has to consider, that the resulting signal from the VMM3a ASIC is about $500-700\,ns$ (see figure~\ref{figsig}), this implies that in order to identify correctly two subsequent events on the same group of electrodes (wires and strips) the signal must be delayed of about least 2 times the signal duration, i.e. $\approx 1\,\mu s$. This corresponds to about $1\,\mathrm{M hits/s}$ or $330\,\mathrm{k events/s}$. 
The theoretical rate limitations in one cassette is then between $330\,\mathrm{k events/s}$ in the worst case and $5\,\mathrm{M events/s}$ in the best case depending on the geographical location of the events.  
\\ In order to improve the data rate, one could connect only one or two hybrids to one FEA and have then rings of three to five nodes. But this would lead to increased cost and a higher complexity of the system. The more convenient option would be to reduce the 128 bit data format for one hit on the FEA to a more compact format. Although it is possible to do optimizations on the hardware and software, in the present configuration the specifications of the hardware and software meet the needs for the Amor instrument in terms of rate.
At Amor the highest achievable rate is about a few MHz in a $10\times10\,cm^2$ area at the detector when measuring the direct beam; this corresponds to about a few hundreds of kHz in a single cassette. On the other hand, when measuring with a sample the rate is about a factor $10^3$ lower than of that of direct beam. In either case, the hardware is capable of coping with these rates.

\subsection{Cooling}
Each VMM3a hybrid consumes about $6\,\mathrm{W}$ of power, this implies that heat management has to be taken into account. Without active cooling the operational temperature of the VMM ASICs rises above $50\,^\circ\mathrm{C}$. The expected VMM3a hybrid lifetime as a function of the operational temperature is reported in~\cite{VMMcool}. According to the ATLAS experiment, the lifetime of a hybrid would be only 5 to 6 years at $45\,^\circ\mathrm{C}$. If the operational temperature decreases by about $10\,^\circ\mathrm{C}$, the expected service life nearly doubles. 
\\ The VMM3a hybrids and the adapters are placed in the electronics box parallel to each other. In case of TBL (or Amor), as shown in figure~\ref{figadapt}, the number of hybrids is 14, i.e. about $84\,\mathrm{W}$. Fans are placed orthogonally at the two ends of the adapter stack (see figures~\ref{fig2} and~\ref{figadapt}). It turns out that the best configuration for lowering the temperature of the ASICs is a push-pull configuration of the fans, two fans pushing air in on one side and two fans pulling air out on the opposite side of the electronics enclosure. With this configuration the average temperature of the hybrids can be kept below $40\,^\circ\mathrm{C}$ which results in an expected lifetime of approximately 8 to 10 years of continuous operation. It is encouraged to switch off the hybrids during long shut downs of the source to extend their lifetime. 
\\ Note that the active cooling is needed to extend the lifetime of the hybrids. Variations in temperature that can alter the gas gain stability of the detector are taken into account by monitoring with a sensor. Pressure, relative humidity and temperature variations of the gas in the detector are monitored with a sensor and this data is used in the post processing to adjust software thresholds to maintain the gas gain stable. 

\section{Commissioning}
After installation at Amor, the detector has been commissioned by exposing it to neutrons. Three types of measurements were carried out to validate the functionality of the detector and its readout chain:
\begin{itemize}
  \item the measurement of the direct beam with and without a Boron-nitrate (BN) mask, 
  \item the measurements of standard samples with well-known features,
  \item the measurements of the direct beam as a function of the opening of the virtual source (VS) to gradually increase the rate at the detector.
\end{itemize}

\subsection{ESS mask and direct beam measurements}
\begin{figure}[!ht]
\centering
\includegraphics[width=14cm,keepaspectratio]{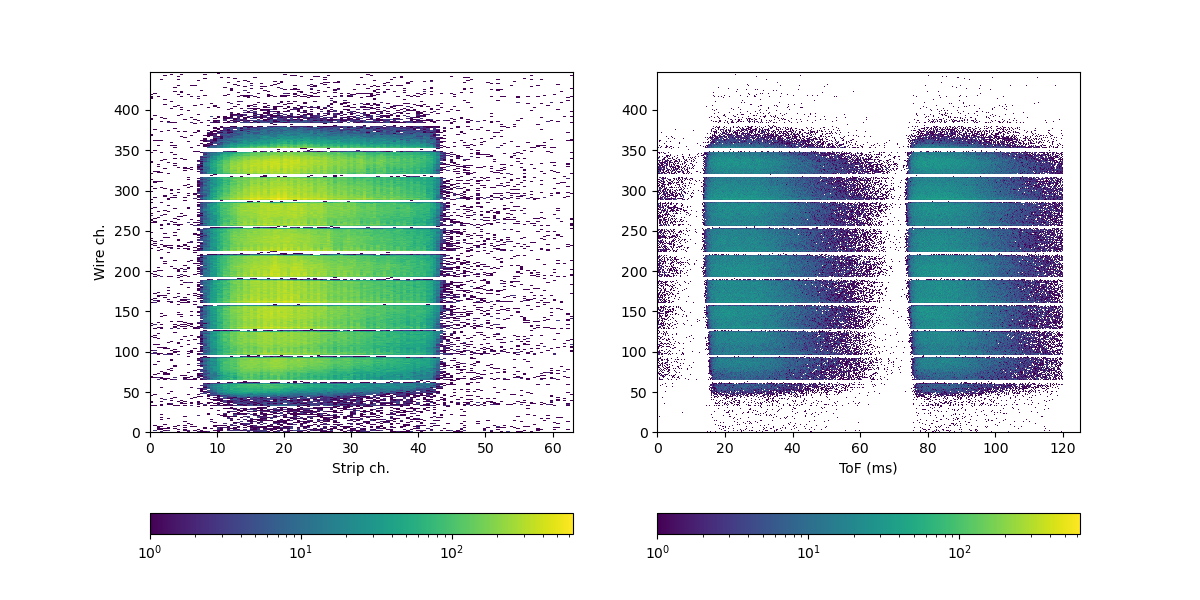}
\caption{\footnotesize Direct beam at the detector. Image of the detector showing the wires versus strips (left) and the ToF histogram showing wires versus ToF (right). The color scale represents counts.}
\label{db1}
\end{figure}

\begin{figure}[!ht]
\centering
\includegraphics[width=10cm,keepaspectratio]{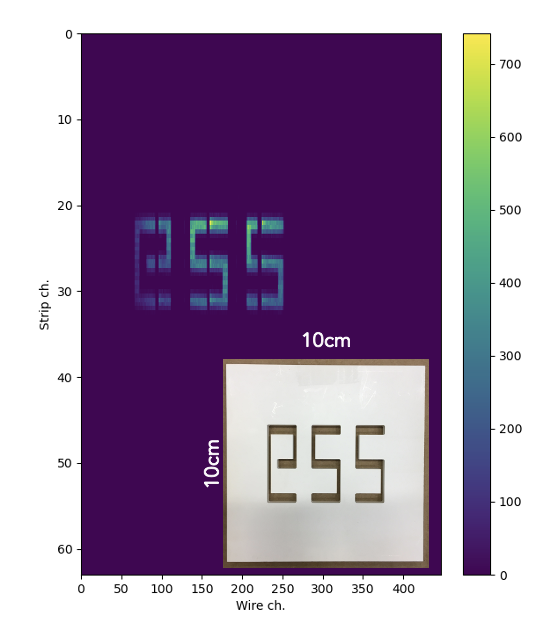}
\caption{\footnotesize A Boron-Nitrate (BN) mask ($10 \times 10\,\mathrm{cm^2}$) with an ESS cut-out (bottom right) to validate the channel mapping of the MB detector and its reconstructed image. The color scale represents counts.}
\label{esspict}
\end{figure}

The detector is at $4\,\mathrm{m}$ from the sample and the beam is focused to the sample position by the Selene guide and diverges from there towards the detector. The beam vertical divergence of 1.6 degrees results in a footprint of about $11\,\mathrm{cm}$ high at the detector. In the horizontal direction the footprint has a size of $12\,\mathrm{cm}$. Figure~\ref{db1} shows the reconstructed image of the beam ($xy$, strips versus wires) and the vertical direction versus ToF histogram ($y$ versus ToF). The chopper at Amor was set to spin with a $120\,\mathrm{ms}$ period. The chopper has two openings with $180^\circ$ phase, hence two neutron pulses are created for each reset signal from the chopper pick-up. This is the reason for the two bunches of neutrons in the ToF histogram in figure~\ref{db1}.
The blades overlap by construction, thus the last bin of each cassette (i.e. wire 31 of each blade) is shown as blank in the image. This shadow, as described in detail in~\cite{MIO_MBAMOR,MIO_MB16CRISP_jinst}, serves as a control for the correct inclination of the overall detector: a larger blank strip indicates that the detector is tilted at a too high angle and an absent shadow indicates that the detector is inclined at a too shallow angle. 
\\In order to verify that the channels and the blades are mapped correctly, a BN mask with a clearly recognisable pattern has been placed in front of the detector window exposed to the direct beam. Figure~\ref{esspict} shows the reconstructed image and the mask itself with the cut-out "ESS".

\subsection{Measurement of samples}
\begin{figure}[!ht]
\centering
\includegraphics[width=14cm,keepaspectratio]{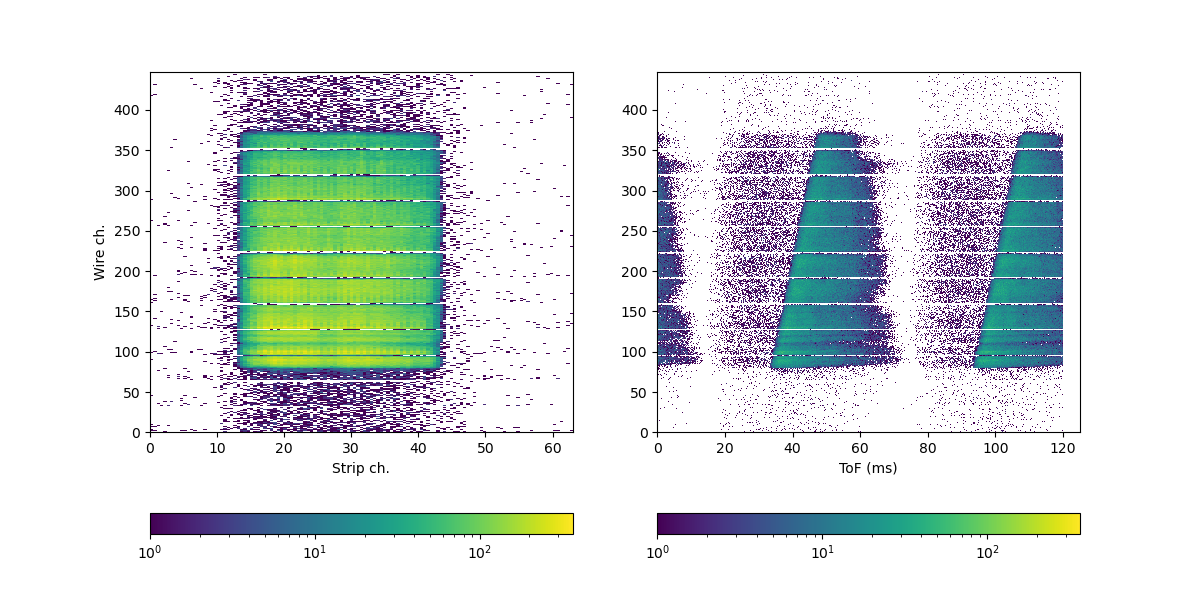}
\caption{\footnotesize Reflection of neutrons from a super-mirror as a sample. Detector image with wires versus strips (left) and the ToF histogram with wires versus ToF (right). The color scale represents counts.}
\label{smpict}
\end{figure}

\begin{figure}[!ht]
\centering
\includegraphics[width=14cm,keepaspectratio]{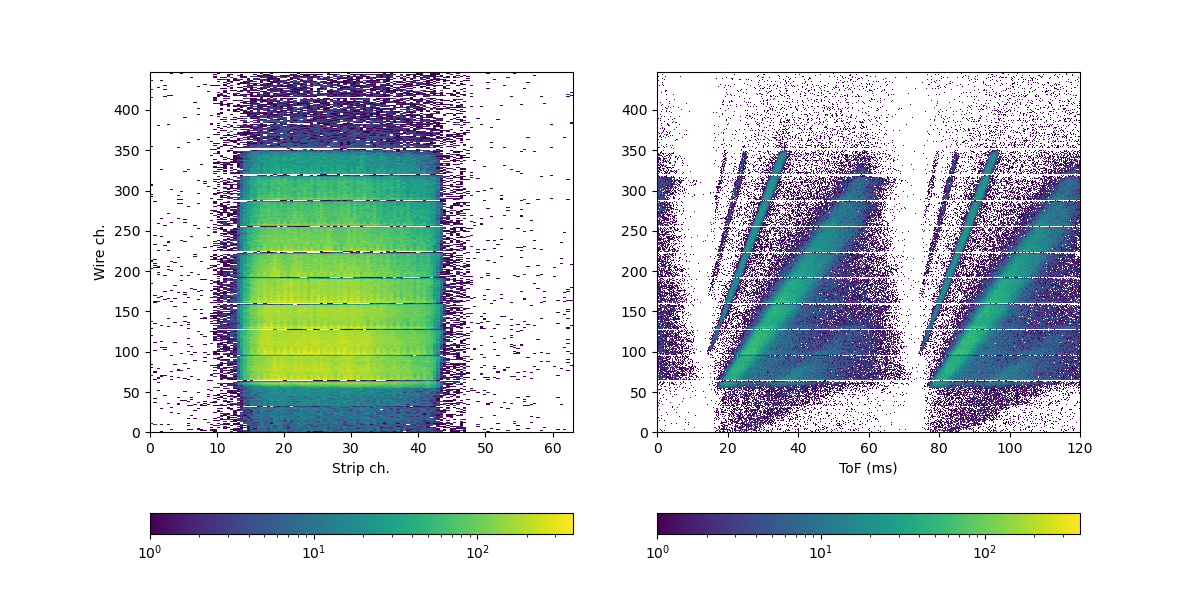}
\caption{\footnotesize Reflection of neutrons from a Ni-Ti sample. Detector image with wires versus strips (left) and the ToF histogram with wires versus ToF (right). The color scale represents counts.}
\label{figniti1}
\end{figure}

\begin{figure}[!ht]
\centering
\includegraphics[width=14cm,keepaspectratio]{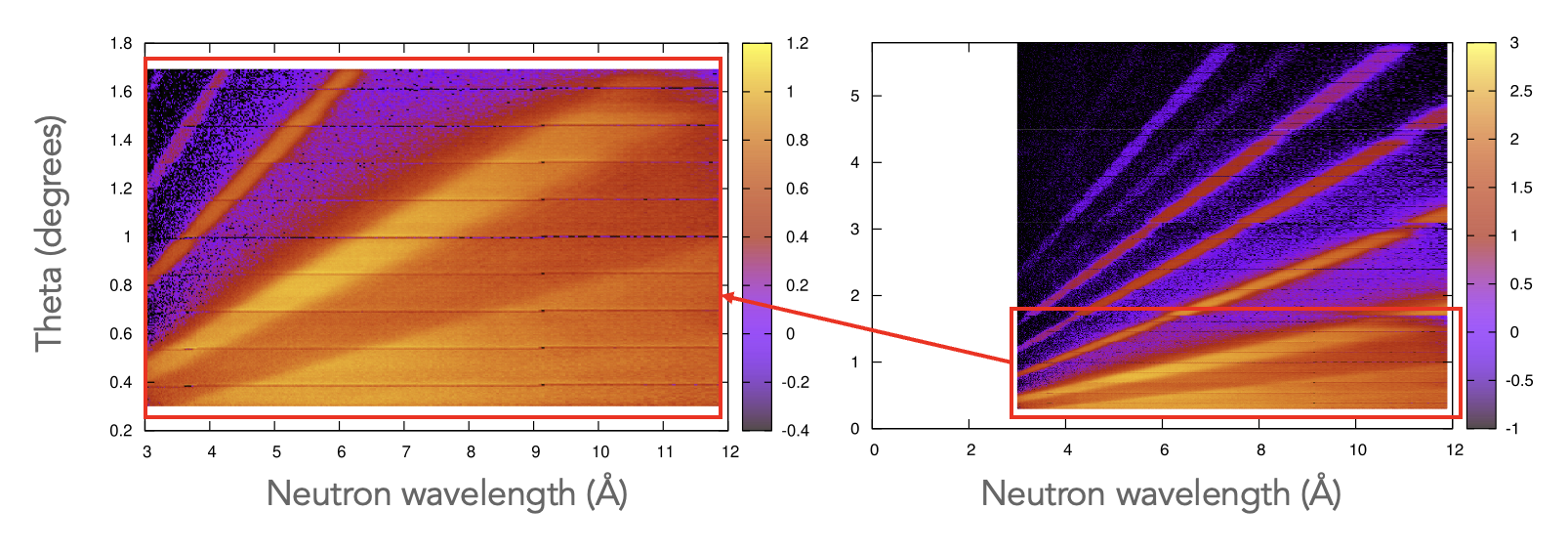}
\caption{\footnotesize Reflection of neutrons from a Ni-Ti multilayer sample after data reduction.  The vertical coordinate of the detector transforms into the final angle $\theta$ and the ToF into the neutron wavelength. 
The single measurement for the lowest sample angle is shown on the left and the combination of four sample angles is shown on the right. The color scale represents counts.}
\label{fignitiall}
\end{figure}

Two samples have been measured: a NiTi super-mirror and a NiTi multi-layer. A NiTi multi-layer is a periodic multi-layer of Ni and Ti. This periodicity leads to Bragg-like diffraction peaks along the momentum transfer ($q_z$). The NiTi super-mirror is a stack of a larger number of Ni and Ti layers than the NiTi multi-layer and the individual thicknesses increase smoothly from the substrate to the surface. This does not result in a Bragg-like diffraction but in an increased reflectivity up to $m \cdot q_z^{critical}(Ni)$; in our case $m=5$. These two samples have very clear features, mainly in the ToF histogram allowing to validate the detector, the readout and the software for reconstruction. The sample is placed on the sample stage, and the detector moved at large angles to intercept the specular reflected beam. The super-mirror, shown in figure~\ref{smpict}, has a very clear edge in ToF as the angle increases. Moving upwards in the blades the edge at which the super-mirror reflects starts at larger ToF (i.e.\ wavelengths). A misplaced blade or mapping would result in a discontinued line in the ToF histogram. 
\\ A NiTi multilayer sample has been measured extensively, being rich of features that can be checked. Figure~\ref{figniti1} shows the reflection from the NiTi multilayer sample at the lowest angle. Measurements were performed at four angles to cover a larger momentum transfer ($q_z$) range. The lines, visible in the ToF plot, are the diffraction order property of the sample. As for the super-mirror they need to align, otherwise a misalignment in the detector channel mapping will affect the intensity. Figure~\ref{fignitiall} shows where the lowest angle is shown on the left and the combination of the four angles on the right. In this picture, the data from physical wire, strip and ToF coordinates have been reduced to the quantities needed to perform the analysis commonly used in neutron reflectometry: theta (the angle of specular reflection) and lambda (the neutron wavelength). These two quantities encode the momentum transfer $q_z=4 \pi \frac{\sin{\theta}}{\lambda}$. 
A simulation of this sample for the purpose of ESTIA has been shown in the ESTIA instrument proposal~\cite{INSTR_ESTIA}.

\subsection{Rate validation and space charge effect}
\begin{figure}[!ht]
\centering
\includegraphics[width=14cm,keepaspectratio]{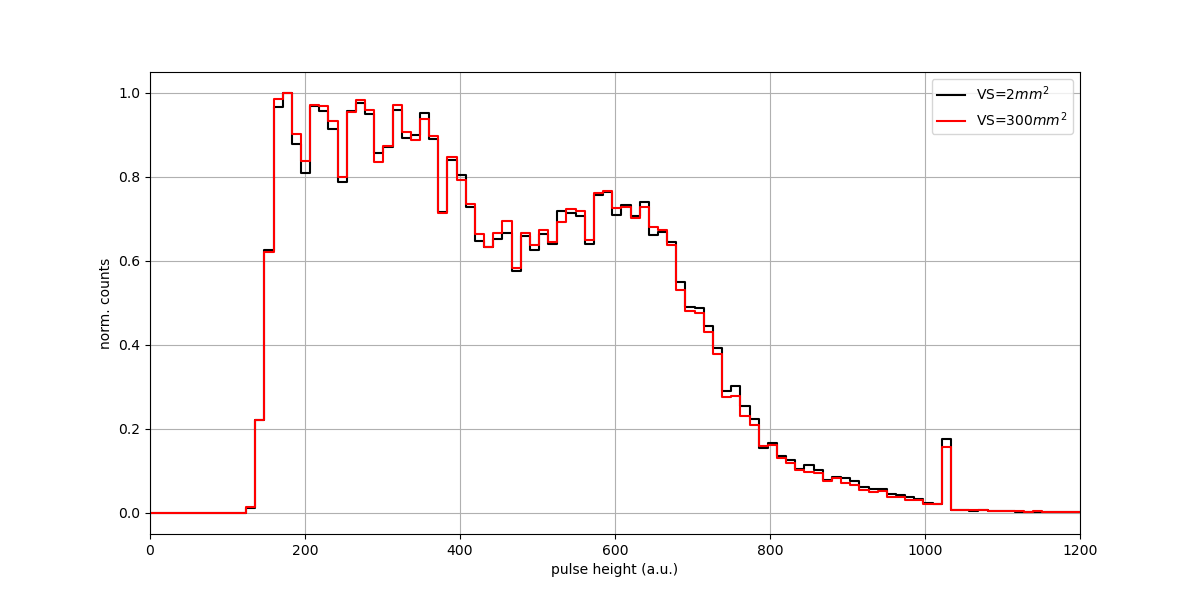}
\caption{\footnotesize Pulse Height Spectrum (PHS) of the 32 wires of a cassette in the direct beam for two different openings of the virtual source. No shift toward lower amplitudes in the PHS at higher rate indicates no saturation due to space charge effects.}
\label{phsfig}
\end{figure}

The virtual source (VS) at Amor is installed before the first Selene guide. By changing the VS dimension the beam intensity is adjusted and the same beam spot is reproduced downstream at the sample position~\cite{INSTR_ESTIA2}. Note that the absolute intensity is not exactly proportional to the area of the source due to the beam inhomogeneities. The validation of the MB detector technology requires evaluating the change in the shape of the pulse height spectrum (PHS) as a sign of the saturation due to the space charge effects. The detector is placed in the direct beam. Figure~\ref{phsfig} shows the PHS of a cassette illuminated by the beam for two openings of the VS: $2\,\mathrm{mm^2}$ and $300\,\mathrm{mm^2}$. The rate in the selected cassette is about $1.5\,\mathrm{kHz}$ and $290\,\mathrm{kHz}$, respectively, for the two openings. The rate in the whole detector area is $15\,\mathrm{kHz}$ and $2200\,\mathrm{kHz}$, respectively.
The two fragments (the alpha particle and the recoiled Lithium-7 nucleus) from the neutron reaction capture in Boron are clearly visible. A contraction of the PHS towards lower energies is not observed, proving the absence of rate saturation induced by space charge effects. The two PHS are normalized, on the vertical axis, to visually facilitate the comparison. The peak at 1023 ADC value of the energy axis is the maximum ADC from the VMM3a ASIC. Note that larger energies are also possible when two, or more, wires (or strips) (multiplicity 2, or more, events) are firing at the same time and the energy of the event is the sum of their two, or more, ADC values. 

\section{Conclusions}
The first of a series of four MB detectors is installed and commissioned at the Amor reflectometer at PSI, including the complete readout electronics chain based on the VMM3a ASIC and ESS readout master module. This experience is invaluable for the future construction, installation and commissioning of three more MB detectors for ESTIA, FREIA and TBL at ESS. The continuous operation at Amor will allow to improve the detector electronics and software while in operation for the users. 

\bibliographystyle{ieeetr}
\bibliography{BIBLIO}
\end{document}